\input harvmac
\input epsf
%
\newbox\hdbox%
\newcount\hdrows%
\newcount\multispancount%
\newcount\ncase%
\newcount\ncols
\newcount\nrows%
\newcount\nspan%
\newcount\ntemp%
\newdimen\hdsize%
\newdimen\newhdsize%
\newdimen\parasize%
\newdimen\spreadwidth%
\newdimen\thicksize%
\newdimen\thinsize%
\newdimen\tablewidth%
\newif\ifcentertables%
\newif\ifendsize%
\newif\iffirstrow%
\newif\iftableinfo%
\newtoks\dbt%
\newtoks\hdtks%
\newtoks\savetks%
\newtoks\tableLETtokens%
\newtoks\tabletokens%
\newtoks\widthspec%
%
%
%
%
\tableinfotrue%
\catcode`\@=11
%
%
\def\tstrut{\vrule height3.1ex depth1.2ex width0pt}%
\def\and{\char`\&}
\def\tablerule{\noalign{\hrule height\thinsize depth0pt}}%
\thicksize=1.5pt
\thinsize=0.6pt
\def\thickrule{\noalign{\hrule height\thicksize depth0pt}}%
\def\ctr#1{\hfil\ #1\hfil}%
%
%
%
%
\tablewidth=-\maxdimen%
\spreadwidth=-\maxdimen%
\def\tabskipglue{0pt plus 1fil minus 1fil}%
%
%
\centertablestrue%
%
%
%
%
\parasize=4in%
\gdef\ARGS{########}
\gdef\headerARGS{####}
\def\@mpersand{&}
{\catcode`\|=13
\gdef\letbarzero{\let|0}
\gdef\letbartab{\def|{&&}}%
\gdef\letvbbar{\let\vb|}%
}
{\catcode`\&=4
\def\ampskip{&\omit\hfil&}
\catcode`\&=13
\let&0
\xdef\letampskip{\def&{\ampskip}}%
\gdef\letnovbamp{\let\novb&\let\tab&}
}
\def\begintable{
   \begingroup%
   \catcode`\|=13\letbartab\letvbbar%
   \catcode`\&=13\letampskip\letnovbamp%
   \def\multispan##1{
      \omit \mscount##1%
      \multiply\mscount\tw@\advance\mscount\m@ne%
      \loop\ifnum\mscount>\@ne \sp@n\repeat%
   }
   \def\|{%
      &\omit\widevline&%
   }%
   \ruledtable
}
\long\def\ruledtable#1\endtable{%
%
%
%
   \offinterlineskip
   \tabskip 0pt
   \def\widevline{\vrule width\thicksize}
   \def\endrow{\@mpersand\omit\hfil\crnorm\@mpersand}%
   \def\crthick{\@mpersand\crnorm\thickrule\@mpersand}%
   \def\crthickneg##1{\@mpersand\crnorm\thickrule
          \noalign{{\skip0=##1\vskip-\skip0}}\@mpersand}%
   \def\crnorule{\@mpersand\crnorm\@mpersand}%
   \def\crnoruleneg##1{\@mpersand\crnorm
          \noalign{{\skip0=##1\vskip-\skip0}}\@mpersand}%
   \let\nr=\crnorule
   \def\endtable{\@mpersand\crnorm\thickrule}%
   \let\crnorm=\cr
%
%
   \edef\cr{\@mpersand\crnorm\tablerule\@mpersand}%
   \def\crneg##1{\@mpersand\crnorm\tablerule
          \noalign{{\skip0=##1\vskip-\skip0}}\@mpersand}%
   \let\ctneg=\crthickneg
   \let\nrneg=\crnoruleneg
   \the\tableLETtokens
%
%
   \tabletokens={&#1}
%
%
   \countROWS\tabletokens\into\nrows%
   \countCOLS\tabletokens\into\ncols%
%
%
   \advance\ncols by -1%
   \divide\ncols by 2%
   \advance\nrows by 1%
%
%
   \iftableinfo %
      \immediate\write16{[Nrows=\the\nrows, Ncols=\the\ncols]}%
   \fi%
%
%
   \ifcentertables
      \ifhmode \par\fi
      \line{
      \hss
   \else %
      \hbox{%
   \fi
      \vbox{%
         \makePREAMBLE{\the\ncols}
         \edef\next{\preamble}
         \let\preamble=\next
         \makeTABLE{\preamble}{\tabletokens}
      }
      \ifcentertables \hss}\else }\fi
   \endgroup
   \tablewidth=-\maxdimen
   \spreadwidth=-\maxdimen
}
\def\makeTABLE#1#2{
   {
   \let\ifmath0
   \let\header0
   \let\multispan0
%
%
   \ncase=0%
   \ifdim\tablewidth>-\maxdimen \ncase=1\fi%
   \ifdim\spreadwidth>-\maxdimen \ncase=2\fi%
   \relax
%
   \ifcase\ncase %
      \widthspec={}%
   \or %
      \widthspec=\expandafter{\expandafter t\expandafter o%
                 \the\tablewidth}%
   \else %
      \widthspec=\expandafter{\expandafter s\expandafter p\expandafter r%
                 \expandafter e\expandafter a\expandafter d%
                 \the\spreadwidth}%
   \fi %
   \xdef\next{
      \halign\the\widthspec{%
      #1
      \noalign{\hrule height\thicksize depth0pt}
      \the#2\endtable
%
      }
   }
   }
   \next
}
\def\makePREAMBLE#1{
   \ncols=#1
   \begingroup
   \let\ARGS=0
   \edef\xtp{\widevline\ARGS\tabskip\tabskipglue%
   &\ctr{\ARGS}\tstrut}
   \advance\ncols by -1
   \loop
      \ifnum\ncols>0 %
      \advance\ncols by -1%
      \edef\xtp{\xtp&\vrule width\thinsize\ARGS&\ctr{\ARGS}}%
   \repeat
   \xdef\preamble{\xtp&\widevline\ARGS\tabskip0pt%
   \crnorm}
   \endgroup
}
\def\countROWS#1\into#2{
   \let\countREGISTER=#2%
   \countREGISTER=0%
   \expandafter\ROWcount\the#1\endcount%
}%
\def\ROWcount{%
   \afterassignment\subROWcount\let\next= %
}%
\def\subROWcount{%
   \ifx\next\endcount %
      \let\next=\relax%
   \else%
      \ncase=0%
      \ifx\next\cr %
         \global\advance\countREGISTER by 1%
         \ncase=0%
      \fi%
      \ifx\next\endrow %
         \global\advance\countREGISTER by 1%
         \ncase=0%
      \fi%
      \ifx\next\crthick %
         \global\advance\countREGISTER by 1%
         \ncase=0%
      \fi%
      \ifx\next\crnorule %
         \global\advance\countREGISTER by 1%
         \ncase=0%
      \fi%
      \ifx\next\crthickneg %
         \global\advance\countREGISTER by 1%
         \ncase=0%
      \fi%
      \ifx\next\crnoruleneg %
         \global\advance\countREGISTER by 1%
         \ncase=0%
      \fi%
      \ifx\next\crneg %
         \global\advance\countREGISTER by 1%
         \ncase=0%
      \fi%
      \ifx\next\header %
         \ncase=1%
      \fi%
      \relax%
      \ifcase\ncase %
         \let\next\ROWcount%
      \or %
         \let\next\argROWskip%
      \else %
      \fi%
   \fi%
   \next%
}
\def\counthdROWS#1\into#2{%
\dvr{10}%
   \let\countREGISTER=#2%
   \countREGISTER=0%
\dvr{11}%
\dvr{13}%
   \expandafter\hdROWcount\the#1\endcount%
\dvr{12}%
}%
\def\hdROWcount{%
   \afterassignment\subhdROWcount\let\next= %
}%
\def\subhdROWcount{%
   \ifx\next\endcount %
      \let\next=\relax%
   \else%
      \ncase=0%
      \ifx\next\cr %
         \global\advance\countREGISTER by 1%
         \ncase=0%
      \fi%
      \ifx\next\endrow %
         \global\advance\countREGISTER by 1%
         \ncase=0%
      \fi%
      \ifx\next\crthick %
         \global\advance\countREGISTER by 1%
         \ncase=0%
      \fi%
      \ifx\next\crnorule %
         \global\advance\countREGISTER by 1%
         \ncase=0%
      \fi%
      \ifx\next\header %
         \ncase=1%
      \fi%
\relax%
      \ifcase\ncase %
         \let\next\hdROWcount%
      \or%
         \let\next\arghdROWskip%
      \else %
      \fi%
   \fi%
   \next%
}%
{\catcode`\|=13\letbartab
\gdef\countCOLS#1\into#2{%
   \let\countREGISTER=#2%
   \global\countREGISTER=0%
   \global\multispancount=0%
   \global\firstrowtrue
   \expandafter\COLcount\the#1\endcount%
   \global\advance\countREGISTER by 3%
   \global\advance\countREGISTER by -\multispancount
}%
\gdef\COLcount{%
   \afterassignment\subCOLcount\let\next= %
}%
{\catcode`\&=13%
\gdef\subCOLcount{%
   \ifx\next\endcount %
      \let\next=\relax%
   \else%
      \ncase=0%
      \iffirstrow
         \ifx\next& %
            \global\advance\countREGISTER by 2%
            \ncase=0%
         \fi%
         \ifx\next\span %
            \global\advance\countREGISTER by 1%
            \ncase=0%
         \fi%
         \ifx\next| %
            \global\advance\countREGISTER by 2%
            \ncase=0%
         \fi
         \ifx\next\|
            \global\advance\countREGISTER by 2%
            \ncase=0%
         \fi
         \ifx\next\multispan
            \ncase=1%
            \global\advance\multispancount by 1%
         \fi
         \ifx\next\header
            \ncase=2%
         \fi
         \ifx\next\cr       \global\firstrowfalse \fi
         \ifx\next\endrow   \global\firstrowfalse \fi
         \ifx\next\crthick  \global\firstrowfalse \fi
         \ifx\next\crnorule \global\firstrowfalse \fi
         \ifx\next\crnoruleneg \global\firstrowfalse \fi
         \ifx\next\crthickneg  \global\firstrowfalse \fi
         \ifx\next\crneg       \global\firstrowfalse \fi
      \fi
\relax
      \ifcase\ncase %
         \let\next\COLcount%
      \or %
         \let\next\spancount%
      \or %
         \let\next\argCOLskip%
      \else %
      \fi %
   \fi%
   \next%
}%
\gdef\argROWskip#1{%
   \let\next\ROWcount \next%
}
\gdef\arghdROWskip#1{%
   \let\next\ROWcount \next%
}
\gdef\argCOLskip#1{%
   \let\next\COLcount \next%
}
}
}
\def\spancount#1{
   \nspan=#1\multiply\nspan by 2\advance\nspan by -1%
   \global\advance \countREGISTER by \nspan
   \let\next\COLcount \next}%
\def\dvr#1{\relax}%
\def\header#1{%
\dvr{1}{\let\cr=\@mpersand%
\hdtks={#1}%
\counthdROWS\hdtks\into\hdrows%
\advance\hdrows by 1%
\ifnum\hdrows=0 \hdrows=1 \fi%
\dvr{5}\makehdPREAMBLE{\the\hdrows}%
\dvr{6}\getHDdimen{#1}%
{\parindent=0pt\hsize=\hdsize{\let\ifmath0%
\xdef\next{\valign{\headerpreamble #1\crnorm}}}\dvr{7}\next\dvr{8}%
}%
}\dvr{2}}
\def\makehdPREAMBLE#1{
\dvr{3}%
\hdrows=#1
{
\let\headerARGS=0%
\let\cr=\crnorm%
\edef\xtp{\vfil\hfil\hbox{\headerARGS}\hfil\vfil}%
\advance\hdrows by -1
\loop
\ifnum\hdrows>0%
\advance\hdrows by -1%
\edef\xtp{\xtp&\vfil\hfil\hbox{\headerARGS}\hfil\vfil}%
\repeat%
\xdef\headerpreamble{\xtp\crcr}%
}
\dvr{4}}
\def\getHDdimen#1{%
\hdsize=0pt%
\getsize#1\cr\end\cr%
}
\def\getsize#1\cr{%
\endsizefalse\savetks={#1}%
\expandafter\lookend\the\savetks\cr%
\relax \ifendsize \let\next\relax \else%
\setbox\hdbox=\hbox{#1}\newhdsize=1.0\wd\hdbox%
\ifdim\newhdsize>\hdsize \hdsize=\newhdsize \fi%
\let\next\getsize \fi%
\next%
}%
\def\lookend{\afterassignment\sublookend\let\looknext= }%
\def\sublookend{\relax%
\ifx\looknext\cr %
\let\looknext\relax \else %
   \relax
   \ifx\looknext\end \global\endsizetrue \fi%
   \let\looknext=\lookend%
    \fi \looknext%
}%
%
%
\def\tablelet#1{%
   \tableLETtokens=\expandafter{\the\tableLETtokens #1}%
}%
\catcode`\@=12
\def \inparg{\leftskip = 35 pt\rightskip = 35pt}
\def \outparg{\leftskip = 0 pt\rightskip = 0pt}
\def\vev#1{\mathopen\langle #1\mathclose\rangle }
\thicksize=0.7pt
\thinsize=0.5pt
\def\ctr#1{\hfil $\,\,\,#1\,\,\,$ \hfil}  
\def\tstrut{\vrule height 2.7ex depth 1.0ex width 0pt}

\def\semi{;\hfil\break}
\def\MSSMnu{\hbox{MSSM}^{\nu}}
\def\frak#1#2{{\textstyle{{#1}\over{#2}}}}
\def\frakk#1#2{{{#1}\over{#2}}}
\def\mbar{{\overline{m}}}

\def\semi{;\hfil\break}
\def\npb{{Nucl.\ Phys.\ }{\bf B}}
\def\prd{{Phys.\ Rev.\ }{\bf D}}
\def\prl{Phys.\ Rev.\ Lett.\ }
\def\plb{{Phys.\ Lett.\ }{\bf B}}
\def\mpla{{Mod.\ Phys.\ Lett.\ }{\bf A}}

\def\thetabar{{\overline \theta}}

\def\tautil{\tilde\tau}
\def\chitil{\tilde\chi}
\def\nutil{\tilde\nu}
\def\mutil{\tilde\mu}

\def\btil{\tilde b}
\def\ctil{\tilde c}
\def\dtil{\tilde d}
\def\etil{\tilde e}

\def\stil{\tilde s}
\def\ttil{\tilde t}
\def\util{\tilde u}

\def\TeV{{\rm TeV}}
\def\GeV{{\rm GeV}}
\def\eV{{\rm eV}}
{\nopagenumbers
\line{\hfil LTH 533}
\line{\hfil hep-ph/0202101}
\line{\hfil Revised Version v4}
\vskip .5in
\centerline{\titlefont Fayet-Iliopoulos $D$-terms, neutrino masses  and }
\centerline{\titlefont anomaly mediated supersymmetry breaking}
\vskip 1in
\centerline{\bf I.~Jack, D.R.T.~Jones and R.~Wild}
\bigskip
\centerline{\it Dept. of Mathematical Sciences,
University of Liverpool, Liverpool L69 3BX, U.K.}
\vskip .3in

We explore, in the context of the MSSM generalised to admit massive 
neutrinos, an extension of  the  Anomaly Mediated Supersymmetry Breaking
solution for the  soft scalar masses to incorporate  Fayet-Iliopoulos
D-terms. The slepton mass problem characteristic of the scenario is
resolved, and the fermion mass  hierarchy is explained via the
Froggatt-Nielsen  mechanism. FCNC problems are evaded by a combination
of universal  doublet charges and Yukawa textures which are diagonalised
by transforming  the left-handed fields only.

\Date{Feb 2002}}

Recently there has been interest in a specific and  predictive 
framework for the origin of soft supersymmetry breaking within the MSSM,
known as Anomaly Mediated Supersymmetry Breaking (AMSB)
\ref\lisa{L. Randall and R. Sundrum, \npb 557 (1999) 79}
\nref\glmr{G.F. Giudice, M.A. Luty, H. Murayama and  R. Rattazzi,
JHEP 9812 (1998) 27}%
\nref\jjpa{I.~Jack, D.R.T.~Jones and A.~Pickering,
\plb426 (1998) 73}%
\nref\kkz{T.~Kobayashi, J.~Kubo and G.~Zoupanos, \plb427 (1998) 291}%
\nref\appp{A. Pomarol and  R. Rattazzi, JHEP 9905 (1999) 013}%
\nref\ggw{T. Gherghetta, G.F. Giudice and J.D. Wells, \npb 559 (1999) 27}%
\nref\clmp{Z. Chacko, M.A. Luty, I. Maksymyk and E. Ponton, 
JHEP 0004 (2000) 001}%
\nref\lura{M.A. Luty and R. Rattazzi, JHEP 9911 (1999) 001}%
\nref\kss{E. Katz, Y. Shadmi and Y. Shirman, JHEP 9908 (1999) 015}%
\nref\jjrg{I.~Jack and D.R.T.~Jones, \plb 465 (1999) 148}%
\nref\jlftm{J.L.~Feng and T.~Moroi, \prd 61 (2000) 095004}%
\nref\gdk{G.D.~Kribs, \prd 62 (2000) 015008}%
\nref\jjfi{I.~Jack and D.R.T~Jones, \plb 482 (2000) 167}%
\nref\shusu{S.~Su, \npb 573 (2000) 87}%
\nref\rzzsw{R. Rattazzi, A. Strumia and J.D. Wells, \npb 576 (2000) 3}%
\nref\jjrp{I.~Jack and D.R.T~Jones, \plb 491 (2000) 151}%
\nref\fepjw{F.E.~Paige and J. Wells, hep-ph/0001249}%
\nref\marcel{M. Carena, K. Huitu and T. Kobayashi, \npb 592 (2000) 164}%
\nref\allanach{B.C. Allanach and A. Dedes, JHEP 0006 (2000) 017}%
\nref\BaerBS{
H.~Baer, J.K.~Mizukoshi and X.~Tata, \plb 488 (2000) 367}%
\nref\GhoshFV{
D.K.~Ghosh, P.~Roy and S.~Roy, JHEP {\bf 0008} (2000) 031}%
\nref\MoroiZB{
T.~Moroi and L.~Randall,
\npb 570 (2000) 455}%
\nref\clppss{Z. Chacko et al, JHEP 0004 (2000) 001}%
\nref\chghro{U. Chattopadhyay, D.K. Ghosh and S. Roy, \prd 62 (2000)  115001}%
\nref\GhoshXP{
D.K.~Ghosh, A.~Kundu, P.~Roy and S.~Roy,
\prd 64 (2001) 115001}%
\nref\nahk{N. Arkani-Hamed et al,
JHEP {\bf 0102} (2001) 041}%
\nref\DattaZW{
A.~Datta and S.~Maity,
\plb 513 (2001) 130}%
\nref\ChackoWQ{
Z.~Chacko et al,  \prd 64 (2001) 055009}%
\nref\DattaER{
A.~Datta, A.~Kundu and A.~Samanta, 
\prd 64  (2001) 095016\semi
E.~Gabrielli, K.~Huitu and S.~Roy, hep-ph/0108246}%
\nref\BMP{E.~Boyda, H.~Murayama and A.~Pierce, hep-ph/0107255}%
\nref\AbeCG{
N.~Abe, T.~Moroi and M.~Yamaguchi, JHEP 0201 (2002) 010}%
\nref\DeCamposWQ{
F.~De Campos et al
\npb 623 (2002) 47}%
\nref\clppst{M.A.~Luty and R. Sundrum, hep-ph/0111231}%
--\ref\clppsu{Z. Chacko and M.A.~Luty, hep-ph/0112172}.
Direct
application of the AMSB solution  to the MSSM leads, unfortunately,  to
negative $(\hbox{mass})^2$ sleptons. 
A number
of possible solutions to this problem have been discussed;  here we
concentrate on a proposal by two of us\jjfi\  (see also \nahk); 
namely the 
introduction of Fayet-Iliopoulos (FI) terms associated with both the  MSSM
$U_1$ and an additional $U'_1$ symmetry. This preserves  the exact RG
invariance of the AMSB solution in rather a minimalist way,  requiring
as it does the introduction of no new fields; the $U'_1$ need not  in
fact be gauged, though the RG invariance requires that we ensure that 
it has vanishing linear mixed anomalies with the MSSM gauge group.  The MSSM
indeed admits two generation-independent, 
mixed-anomaly-free  $U_1$ groups, the existing $U_1^Y$ and another
(which could be chosen to be  $U_1^{B-L}$\nahk, or a linear
combination of it and $U_1^Y$).

Extension of this scenario to include massive neutrinos meets an 
obstacle inasmuch as there is no flavour independent global
$U'_1$ symmetry possible for a superpotential incorporating both 
neutrino Yukawa couplings and Majorana masses for right-handed
neutrinos.  We are therefore driven to consider flavour dependent
symmetries\foot{for some alternative ideas see Ref.~\nahk},  
and choose to make a virtue out of necessity in this regard
by re-examining  the well-travelled path of Yukawa coupling textures
associated with  a $U_1$ symmetry
\ref\cdfr{C.D. Froggatt and H.B. Nielsen, \npb 147 (1979) 277\semi
P. Ramond, R.G. Roberts and  G.G. Ross,
 \npb 406 (1993) 19\semi
M.~Leurer, Y.~Nir and N.~Seiberg,  \npb 398 (1993) 319\semi
L.E.~Ibanez and G.G.~Ross, \plb 332 (1994) 100\semi
P.~Binetruy and P.~Ramond, \plb 350 (1995)\semi  
E. Dudas, S. Pokorski and C.A. Savoy, \plb 356 (1995) 45\semi
J.K.~Elwood, N.~Irges and P.~Ramond,   \prl  81 (1998)  5064\semi
S.F.~King, \npb 562 (1999) 57\semi
S.~Tanaka,  \plb  480 (2000)  296 \semi
S.~Lola and G.G.~Ross,
\npb 553 (1999) 81\semi
M.S.~Berger and K.~Siyeon, \prd  63 (2001) 057302 ; ibid 
D64 (2001)  053006}. 
Most of the  (considerable) literature on this 
subject has dealt with  an {\it anomalous\/} $U'_1$  (with anomaly
cancellation finally achieved via the Green-Schwarz  mechanism
\ref\grsch{M.~Green and J.~Schwarz, \plb 149 (1984) 117}). This
route is not open to us, however, as we require cancellation
of the mixed gauge-$U'_1$ anomalies.  We believe, however, that the
conclusion  that texture generation via an anomaly-free $U'_1$ is
impossible is  a consequence of assumptions which, while plausible, are
not strictly  necessary (what constitutes necessity being in these
matters,  we admit,  a question of taste). 

Thus our goal is to show that the MSSM with massive neutrinos 
(which we will call the $\hbox{MSSM}^{\nu}$) admits a global $U'_1$
which enables us to solve the AMSB tachyonic slepton problem while 
simultaneously reproducing the well-known hierarchies
\ref\rgrss{ 
P.M.~Fishbane and P.Q.~Hung,
\prd 45 (1992)  293\semi
Y.~Koide, H.~Fusaoka and C.~Habe,
\prd 46 (1992)  4813\semi
A.A.~Maslikov and G.G.~Volkov,
ENSLAPP-A-382-92\semi
P.~Ramond, R.G.~Roberts and G.G.~Ross, Ref.~\cdfr
}
\eqn\massrats{
m_{\tau}:m_{\mu}:m_e = m_b:m_s:m_d = 1:\lambda^2:\lambda^4,
\quad\hbox{and}\quad
m_t:m_c:m_u = 1:\lambda^4:\lambda^8,
}
where $\lambda \approx 0.22$. 

We will assume that Yukawa terms are generated via the Froggatt-Nielsen (FN) 
mechanism\cdfr: specifically, from 
higher dimension terms involving $\hbox{MSSM}^{\nu}$ singlet fields 
$\theta_{t,b,\tau}$ with each $\theta$ associated with 
a {\it particular\/}  Yukawa matrix, via terms such as 
$H_2 Q_i t^c_j (\frakk{\theta_t}{M_U})^{a_{ij}}$, where $M_U$ represents 
the scale of new physics.  
Then if we require Yukawa textures consistent with the above hierarchies,
and also require that the mixed anomalies 
cancel, we are in general led to consider different charges for each of 
$\theta_{t,b,\tau}$. If we choose $\theta$-charges
\eqn\spchgs{
q_{\theta_t} = -1, \quad q_{\theta_b} = 2-\frak{\Delta}{2}, 
\quad q_{\theta_{\tau}} = 2 -\frak{\Delta}{3} ,
}
where $\Delta = h_1 + h_2$,
and the charge assignments shown in Table~1,
\vskip3em
\vbox{
\begintable
 Q_i| t^c_2 | t^c_3  | b^c_1 | b^c_2  |b^c_3 
\cr
8-t^c_1-h_2
|t^c_1-4  | t^c_1-8|3h_2+h_1+t^c_1-16|2h_2+t^c_1-12 |t^c_1-h_1+h_2-8
\endtable
\bigskip
\begintable
L_i |  \tau^c_1 | \tau^c_2 | \tau^c_3 
\cr
3t^c_1-\frak{1}{3}h_1+\frak{8}{3}h_2-24|\frak{2}{3}h_1
-\frak{4}{3}h_2+16-3t^c_1 |20-2h_2-3t^c_1 
|24-\frak{2}{3}h_1-\frak{8}{3}h_2-3t^c_1 
\endtable
\inparg
{\noindent\hskip 2cm {\it Table~1:\/} The $U'_1$-charges}
\bigskip \outparg}
\noindent then the mixed anomalies cancel and we find textures given by 
\eqn\texts{
Y_t \sim \pmatrix{\lambda^8&\lambda^4&1\cr
\lambda^8&\lambda^4&1\cr\lambda^8&\lambda^4&1}, \quad
Y_b \sim \pmatrix{\lambda^4&\lambda^2&1\cr
\lambda^4&\lambda^2&1\cr\lambda^4&\lambda^2&1\cr}, \quad
Y_{\tau} \sim \pmatrix{\lambda^4&\lambda^2&1\cr
\lambda^4&\lambda^2&1\cr\lambda^4&\lambda^2&1},}
where we assume  
$\vev{\theta_{t,b,\tau}}/M_U\approx\lambda\approx 0.22$. 
The powers of $\lambda$ are determined by relations such 
as $h_1 + Q_i + b^c_1 + 4q_{\theta_{b}} = 0$. The equality of the rows in 
each matrix corresponds to generation independent 
{\it \/} doublet charges $Q_i$ and 
$L_i$. (We use the same notation for the field and its $U_1'$ charge; it
should be clear from the context which is intended.)

Textures of this form have in fact been considered before 
in the context of D-branes
\ref\ekk{L.~Everett, G.L.~Kane and S.F.~King, JHEP 0008 (2000) 012}, and 
termed ``single right-handed democracy''.
It is easy to show that the eigenvalues of 
$Y_{t,b,\tau}$ above lead to the mass textures of Eq.~\massrats. 
Another feature of textures of this generic form is that since to a 
good approximation we have 
\eqn\ysqrd{
Y_{t}^T Y_{t} \sim \pmatrix{0&0&0\cr 0&0&0\cr 0&0&1\cr}}
(similarly for $Y_{b,\tau}$) it is evident  
that the rotation on the RH fields required to diagonalise 
the mass matrix will be of  the generic form
\eqn\rightrot{
U_R \sim \pmatrix{\cos\alpha& \sin\alpha&0\cr 
-\sin\alpha&\cos\alpha&0\cr 0&0&1\cr}.}
Moreover, by considering the specific textures shown in Eq.~\texts\ 
in the approximation that we set to zero the first column of each matrix 
it is easy to show that in our case $U_R$ will always be 
close to the unit matrix\ekk. This 
will be significant later when we consider flavour changing neutral 
currents (FCNCs). 
If we assume the specific forms
\eqn\qyuks{
Y_t \propto \pmatrix{a_t\lambda^8&d_t\lambda^4&1 + O(\lambda^2)\cr
b_t\lambda^8&e_t\lambda^4&1 + O(\lambda^2)\cr
c_t\lambda^8&f_t\lambda^4&1 + O(\lambda^2)}
\quad\hbox{and}\quad 
Y_b \propto \pmatrix{a_b\lambda^4&d_b\lambda^2&1 + O(\lambda^2)\cr
b_b\lambda^4&e_b\lambda^2&1 
+ O(\lambda^2)\cr c_b\lambda^4&f_b\lambda^2&1 + O(\lambda^2)\cr}}
then we obtain for the CKM matrix the texture
\eqn\ckmtext{CKM \sim  \pmatrix{1&1&\lambda^2\cr
1&1&\lambda^2\cr\lambda^2&\lambda^2&1}
}
which is not of the form of the standard Wolfenstein parametrisation, 
\eqn\ckmwolf{CKM_W  \sim \pmatrix{1&\lambda&\lambda^3\cr
\lambda&1&\lambda^2\cr \lambda^3&\lambda^2&1}
}
It does, however,  reproduce the most significant feature, which is 
the smallness 
of the couplings to the third generation.
\foot{We disagree somewhat with Ref.~\ekk, where it is asserted 
that the Wolfenstein  texture follows if we replace the $(13)$ elements 
of both $Y_t$ and $Y_b$ in Eq.~\qyuks\ by $1+O(\lambda)$. With this 
particular form for $Y_{t,b}$ it is straightforward to establish 
(by either numerical or analytic means) that the CKM matrix 
would have a texture similar to Eq.~\ckmtext\ but with $\lambda^2$ 
replaced everywhere by $\lambda$.}   
It follows that it is possible to exhibit 
explicit forms of $Y_t,Y_b$ with ``reasonable'' coefficients 
reproducing the observed CKM matrix,  for example:
\eqn\qyuksnum{
Y_t \propto \pmatrix{
   11.35 \lambda^8  & 0.915\lambda^4  & 1.048\cr
         1.244 \lambda^8  & 3.336\lambda^4  & 0.970\cr
         3.362 \lambda^8  & -4.266 \lambda^4  & 0.980\cr}
\quad   Y_b \propto \pmatrix{
0.487 \lambda^4  & 0.281 \lambda^2  & 1.063\cr
         -1.311\lambda^4  & 0.398 \lambda^2  & 1.008\cr
         -0.514 \lambda^4  & -0.750 \lambda^2  & 0.925\cr}.}
The $ub$ and $td$ entries in the CKM matrix 
are comparatively sensitive to changes in the 
coefficients in Eq.~\qyuksnum, because our texture form, Eq.~\qyuks, does 
not {\it naturally\/} explain the factor of $10$ difference between 
these entries and the $cb$ and $ts$ ones.

The charges shown in Table~1 have been chosen to provide cancellation 
of the mixed $U'_1 (SU_3)^2$, $U'_1 (SU_2)^2$, and $U'_1 (U_1)^2$ 
anomalies. This is what is required to render our scalar mass solution 
RG invariant; but it is also of interest to examine the remaining 
anomalies involving $U'_1$. The $(U'_1)^2 U_1$, $(U'_1)^3$ and 
$(U'_1)$-gravitational  anomalies  are proportional respectively to
\eqn\anomquad{
A_Q = \Delta\left[8h_2+6t^c_1-\frak{224}{3}+\frak{14}{9}\Delta\right],
}

\eqn\anomcub{\eqalign{A_C  &= 
-108 t^c_1 \Delta h_2 
- 2368 \Delta - \frak{632}{3} \Delta^2  + 640 \Delta h_2
+ 816 t^c_1 \Delta + \frak{92}{3} h_2 \Delta^2  - 48 \Delta h_2^2\cr
&+ 16 t^c_1 \Delta^2   -54 (t^c_1)^2  \Delta + \frak{16}{9}\Delta^3\cr
&+3(9(t^c_1)^2-168t^c_1+24h_2t^c_1+16h_2^2+816-224h_2)(3t^c_1+4h_2-28),\cr}}
and
\eqn\anomgrav{
A_G = 3(3t^c_1+4h_2-28).
}
Note that there will be additional contributions to $A_G$ and $A_C$ from 
any $\MSSMnu$ singlet fields with $U'_1$ charges, such as, of course,
the $\theta$-fields introduced above,  unless they are accompanied by 
oppositely charged partners. In the specific case of the $\theta$-fields, 
such $\thetabar$-partners (if they exist) 
must be forbidden from generating textures in order to 
preserve the patterns of Eq.~\texts. 
One  reason for assuming the $\thetabar$s
exist is that unless they do (and have vevs approximately equal to the 
corresponding $\theta s$)
then the quadratic $D$-terms for the $U'_1$ (if it is gauged) 
will generate 
large masses for all the $\hbox{MSSM}^{\nu}$ fields
\ref\ibross{L.E. Ib\'a\~nez and G.G. Ross, \plb 332 (1994) 100}. 
The possible generation of weak-scale contributions 
to the sparticle masses via this route and their impact 
on flavour-changing processes was discussed recently in 
Ref.~\ref\MurakamiHK{
B.~Murakami, K.~Tobe and J.D.~Wells,
\plb 526 (2002) 157}.

For our purpose it is not necessary to gauge $U'_1$, or remove 
its anomalies other than the linear mixed ones; let us, however, 
explore (en passant) the option of rendering it completely anomaly-free.   
It is easy to show that the conditions 
\eqn\anomfree{\Delta = 0\quad \hbox{and }\quad 3t^c_1+4h_2-28 = 0} 
are necessary and sufficient to give $A_Q = A_C = A_G = 0$ above. 
(Cancellation of all the anomalies requires the existence of  
the $\theta$-partners $\thetabar$ as described above;
if the $\theta$-contributions in $A_C$ and $A_G$ are not cancelled,
it is straightforward to show there is no
charge assignment such that $A_Q = A_C = A_G = 0$.)
The solution Eq.~\anomfree\ may be of interest in the non-AMSB context, 
providing as it does a fully anomaly free $U'_1$ (and one which would remain 
anomaly-free with the introduction of $SU_3\otimes SU_2\otimes U_1$ singlets 
in $\pm$   pairs). However it turns out that 
we cannot with these conditions resolve the slepton mass problem, as 
we shall now show. 

Including the FI contributions the scalar masses are given by\jjfi
\eqn\mssrel{\eqalign{
\mbar^2_{Q} &= m^2_{Q} +\frak{1}{6}\zeta_1 +\zeta_2 Q_i\delta^i_j,\quad
\mbar^2_{t^c} = m^2_{t^c} -\frak{2}{3}\zeta_1 + \zeta_2 t^c_i\delta^i_j,\cr
\mbar^2_{b^c} &= m^2_{b^c} +\frak{1}{3}\zeta_1  +\zeta_2 b^c_i\delta^i_j,\quad
\mbar^2_L =  m^2_L -\frak{1}{2}\zeta_1 + \zeta_2 L_i\delta^i_j,\cr
\mbar^2_{\tau^c} &= m^2_{\tau^c} +\zeta_1 + \zeta_2 \tau^c_i\delta^i_j,
\quad \mbar^2_{H_1} =  m^2_{H_1} -\frak{1}{2}\zeta_1 + \zeta_2 h_1,\cr
\mbar^2_{H_2} &=  m^2_{H_2} +\frak{1}{2}\zeta_1 + \zeta_2 h_2\cr}}
where $\zeta_{1,2}$ are constants and where
$m^2_{Q},\cdots$ denote the AMSB contributions. For example\lisa-\jjpa,
\eqn\mamsb{(m^2_{Q})^i{}_j 
= \frak{1}{2}|m_0|^2\mu\frakk{d(\gamma_Q)^i{}_j}{d\mu},}
where $\gamma_Q$  is the quark doublet anomalous dimension matrix and 
$m_0$ is the gravitino mass. The slepton mass problem is the fact that 
$m^2_L$ and $m^2_{\tau^c}$ have negative eigenvalues. 
However, as we shall see, we can
choose $U_1'$ charges so that for some region of $\zeta_{1,2}$ parameter
space, the eigenvalues of $\mbar^2_L$ and $\mbar^2_{\tau^c}$ are all 
positive, and indeed we obtain a fully realistic mass spectrum.

In fact, it is easily 
shown that with the charge assignments shown in Table~1, and $\Delta = 0$, 
there exists some range of $\zeta_{1,2}$ leading 
to positive  FI contributions for both $\mbar^2_{\tau^c}$ and $\mbar^2_L$ 
if and only if 
\eqn\posi{3t^c_1+4h_2 < 24 \quad\hbox{or}\quad 3t^c_1 + 4h_2 > 32,}
so that the fully anomaly-free solution Eq.~\anomfree\ is excluded. 
We will nevertheless choose to impose cancellation of the quadratic
($A_Q$)  anomaly since (unlike $A_C$ and $A_G$) it cannot be affected by
a  $SU_3\otimes SU_2\otimes U_1$ singlet sector.  Obviously for $A_Q=0$
we require either $\Delta=0$ or 
$8h_2+6t^c_1-\frak{224}{3}+\frak{14}{9}\Delta=0$. We start  with the
$\Delta=0$ case, postponing discussion of the second possibility  until
later. On the one hand this means that a Higgs $\mu$-term is allowed, 
and so we have no solution to the $\mu$-problem; on the other hand from 
Eq.~\spchgs\ we see that the same $\theta$-field will in fact serve for both
down and  charged lepton matrices. Moreover we do not need to forbid
terms of the kind, for  example, $H_2 Q_i t^c_j
(\frakk{\theta_b}{M_U})^{a_{ij}}$, since no such  (gauge invariant) term
can be constructed. 
 
We turn now to the issue of  neutrino masses.
If we wish to generate them via the seesaw 
mechanism then this suggests that we introduce 
three right-handed neutrinos with zero $U'_1$ charges; 
however to obtain Dirac mass terms of the 
form $H_2 L\nu^c$ we  then require a $\theta$-field capable of being 
matched to $L_i + h_2 = 3t^c_1+4h_2 - 24$.
But if we examine potential dimension 4 $R$-parity violating
operators, we  find that the possible superpotential operators 
 $t^c_2 b^c_2 b^c_3$, $t^c_1 b^c_1
b^c_3$, $Q_i L_j b^c_3$ and $L_i L_j \tau^c_3$ have the same charge as $L_i
h_2$,  with disastrous consequences for proton decay, if 
we allow them to be generated at the same level as the  $H_2 L\nu^c$ terms. 
We could choose  to impose $R$-parity conservation, but an attractive
alternative is to  introduce only two right-handed neutrinos, 
with charges $\nu^c_{1,2} = \pm q_{\nu}$  and 
introduce $\theta_{\nu}$ with charge $q_{\theta_{\nu}}$ such that 
\eqn\nuspur{
L_i + h_2 + q_{\nu}+nq_{\theta_{\nu}}=0, \quad\hbox{and}\quad
L_i + h_2 - q_{\nu} +mq_{\theta_{\nu}}=0}
for integer $m,n$. 
It is easy to show that if we choose, for example, $n = 2$ and $m = 1$  
and $ q_{\theta_{\nu}}= -9$, then  none of the 
$R$-parity violating Yukawas mentioned above (nor any $R$-violating 
Yukawa) 
can be generated using the 
available $\theta$-charges. Consequently unrealistic proton decay 
is prevented. 
We obtain a $\nu^c$ matrix of the form (for consideration of 
various forms for the $\nu^c$ matrix see for example 
\ref\KingMB{ S.F.~King, \npb 576 (2000)  85}) 

\eqn\majmat{ M_{\nu^c} = \pmatrix{0&M^{\nu}_1\cr M^{\nu}_1&0}} 
which
has non-zero determinant and therefore will serve  for the seesaw.
\foot{With the choice of $n,m,q_{\theta_{\nu}}$ above, 
there would be the possibility of a $O(\lambda M)$ entry in place 
of the zero for $(M_{\nu^c})_{22}$; this 
does not change any of our conclusions   in an essential way (in particular 
the matrix $m_{\nu}$ below retains a zero eigenvalue).}  
Moreover the Dirac matrix from $H_2 L \nu^c$ takes  the form
\eqn\mdirac{m_D
 = \pmatrix{a_{\nu}\lambda^n&d_{\nu}\lambda^m\cr
b_{\nu}\lambda^n&e_{\nu}\lambda^m\cr c_{\nu}\lambda^n&f_{\nu}\lambda^m\cr}v_2,}
(where $\vev{H_2}=\pmatrix{0\cr v_2\cr}$) 
and the eigenvalues of the resulting LH neutrino mass matrix 
\eqn\numasses{
m_{\nu} = m_D M_{\nu^c}^{-1} (m_D)^T}
are given by  
\eqn\nueigs{
m_{\nu_{1..3}} = 0, (M^{\nu}_1)^{-1}\lambda^{m+n}(n_1 \mp \sqrt{n_2})}
where 
\eqn\nueigsb{\eqalign{
n_1 &= a_{\nu}d_{\nu} +b_{\nu}e_{\nu} +c_{\nu}f_{\nu},\cr
n_2 &= d_{\nu}^2 a_{\nu}^2+e_{\nu}^2 b_{\nu}^2+
f_{\nu}^2 c_{\nu}^2+e_{\nu}^2 c_{\nu}^2+b_{\nu}^2 f_{\nu}^2
+d_{\nu}^2 b_{\nu}^2+a_{\nu}^2 e_{\nu}^2+d_{\nu}^2 c_{\nu}^2
+a_{\nu}^2 f_{\nu}^2.\cr}}
It is clear that $|m_{\nu_3}| > |m_{\nu_2}|$ and although 
both are the same order in 
$\lambda$ the relation $|m_{\nu_3}| = 10 |m_{\nu_2}|$ 
holds for ``reasonable values'' 
of $a_{\nu}\cdots f_{\nu}$. Such a  hierarchy accommodates 
solar and atmospheric neutrino data-see for example Ref.~\KingMB. 
With masses\foot{The recently reported measurement
\ref\Klapdor-KleingrothausKE{ H.V.~Klapdor-Kleingrothaus, 
{\it et al}, \mpla 16 (2001)  2409}\ in neutrinoless double beta decay   
of a Majorana neutrino mass  
in the region $0.11-0.56\eV$ is not readily accommodated within our scenario; 
we note, however, some 
controversy\ref\aalseth{
F.~Feruglio, A.~Strumia and F.~Vissani, hep-ph/0201291\semi
C.E.~Aalseth {\it et al.}, hep-ex/0202018\semi
H.V.~Klapdor-Kleingrothaus, hep-ph/0205228\semi
H.L.~Harney, hep-ph/0205293
} 
regarding this result.}
\eqn\numsp{
m_{\nu_1} = 0,\quad
m_{\nu_2} = 5\times 10^{-3}\eV,\quad
m_{\nu_3} = 5\times 10^{-2}\eV}
we would also expect large mixing angles $\theta^{\nu}_{12},\theta^{\nu}_{23}$
and a small mixing angle $\theta^{\nu}_{13}$ in the rotation to 
the neutrino mass eigenstate basis from the charged lepton 
mass eigenstate basis. It is easy 
to construct examples (without fine-tuning) that give rise to precisely 
this structure within our scenario.
In order to generate a neutrino spectrum in the region of Eq.~\numsp, 
we would require (assuming $a_{\nu}\cdots f_{\nu}$ are $O(1)$) 
\eqn\majres{
M^{\nu}_1 \sim \lambda^{(m+n)}10^{16}\GeV,}
or $M^{\nu}_1\sim 10^{14}\GeV$ in the case $m+n = 3$.
An example consistent with our requirements is given by:
\eqn\lepmat{
m_L = \pmatrix{-0.56\lambda^4 &0.56\lambda^2&-1.07\cr
-0.36\lambda^4&-2.11\lambda^2&-0.22\cr
-0.49\lambda^4&-0.14\lambda^2&-1.43\cr},\quad
m_D
 = \pmatrix{\lambda^2&\lambda\cr
2\lambda^2&5\lambda\cr 3\lambda^2&1.9\lambda\cr}v_2,}
(where $m_L$ is the charged lepton mass matrix) 
when we obtain the neutrino spectrum of Eq.~\numsp\ for 
$M^{\nu}_1 \sim 2.4\times10^{14}\,\GeV$, with 
$\theta^{\nu}_{12} = 0.53,$ $\theta^{\nu}_{23} = 0.78$ and 
$\theta^{\nu}_{13} = 0$. Of course the result for $M_1^{\nu}$ is sensitive 
to the overall scale of $m_D$.

We have therefore achieved our goal of incorporating neutrino masses
into the  AMSB paradigm; with the added bonus that no additional
symmetries  (beyond $U'_1$) are required to adequately suppress proton
decay. The chief feature distinguishing the sparticle spectrum 
from the massless  neutrino case considered in Ref.~\jjfi\ is the
splitting between  the first and second generation right-handed squarks
and sleptons (the degeneracy persists in the LH case because  of the
generation independent doublet $U'_1$ charges).

For example, with $\tan\beta = 5$, 
gravitino mass $m_0 = 40\TeV$, $\zeta_1 = -0.02\TeV^2$ and 
$\zeta_2 = 0.0227\TeV^2$, $h_2 =12$, $t_1^c = -7/2$,
we find $|\mu| = 571\GeV$, and choosing $\hbox{sign}\,\mu = -1$ 
we obtain the following spectrum: 
\eqn\spect{\eqalign{
m_{\ttil_1} & =869, \quad m_{\ttil_2}=484, \quad m_{\btil_1}=825, \quad 
m_{\btil_2}=1082, 
\quad m_{\tautil_1}=148,\cr
\quad m_{\tautil_2} &= 442,\quad
m_{\util_L,\ctil_L} =931, \quad m_{\util_R}=908, \quad m_{\ctil_R}=856
\quad m_{\dtil_L, \stil_L} =934,\cr
\quad m_{\dtil_R} & = 998, \quad m_{\stil_R} = 1042,
\quad m_{\etil_L, \mutil_L} =149,\quad m_{\etil_R}=117,\cr
\quad m_{\mutil_R} & = 323\quad 
m_{\nutil_e, \nutil_{\mu}} = 126,\quad m_{\nutil_{\tau}}=125\quad 
m_{h,H} = 122, 166,\cr 
m_A &= 161, \quad  m_{H^{\pm}} = 181,
\quad m_{\tilde\chi_{1,2}^{\pm}} = 112, 575,\cr
m_{\chitil_{1\cdots4}} & =  111, 369, 579, 579 \quad m_{\tilde g} = 1007,\cr}}
where all masses are given in GeV. The squarks $\ttil_1, \btil_1$ and
$\tautil_1$ couple more strongly to $t_L$, $b_L$ and $\tau_L$ respectively, 
though (for our chosen $\tan\beta$) the $\ttil_{1,2}$ mixing is of course 
substantial.  
For a given $m_0$, an acceptable vacuum is obtainable for only a  very
restricted range of the parameters $\tan\beta, \zeta_1,\zeta_2$.  
The main constraints on the boundary of the allowed region 
come from the slepton and Higgs masses.
The triangular region in the
$\zeta_{1,2}$ plane which meets these constraints is shown
in Fig.~1  (for $m_0=40\hbox{TeV}$, and $\tan\beta=5$).
The LSP can be the neutral wino (as in Eq.~\spect\ above), 
or the $\nutil_{\tau}$; note, however, 
that there is a small region in which the 
$\tilde e_R$ is the LSP, 
which we would exclude on cosmological grounds.  
The point corresponding to our example of 
Eq.~\spect\ is indicated by an asterisk in the diagram. 

\smallskip
\epsfysize= 2.8in
\centerline{\epsfbox{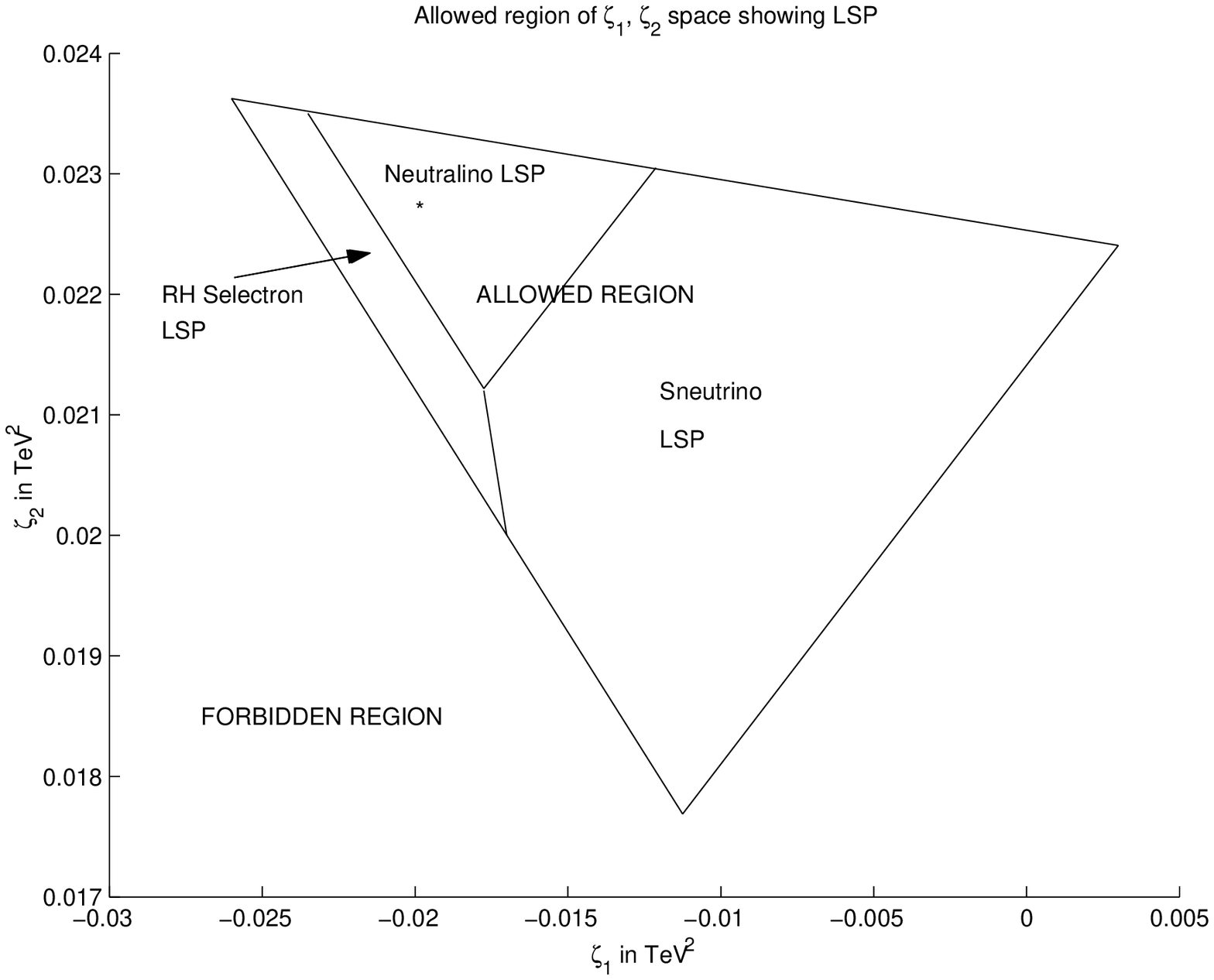}}
\inparg
{\it \noindent Fig.1:
Allowed values of $\zeta_{1,2}$ for $\tan \beta=5$,
$m_0=40\TeV$ and $\hbox{sign}\,\mu = -1$.
}
\smallskip  
\outparg

For discussion of sneutrino cold dark matter, 
see Refs.~\ref\snucdm{L.J.~Hall, T.~Moroi and H.~Murayama,
\plb 424 (1998) 305\semi
S.~Kolb {\it et al}, \plb 478 (2000) 262}. 
For $m_0\sim 40\to50\TeV$ we find 
a maximum possible value of $\tan\beta\sim 15 $. With, for example,  
$m_0 = 50\TeV$ and $\tan\beta = 10$  we find for 
$\zeta_1 = -0.03, \zeta_2 = 0.032$ a spectrum similar to the above with 
generally increased masses. 
The spectrum always features the near-degenerate wino triplet that is
characteristic of the AMSB scenario. However when  
the LSP is the $\nutil_{\tau}$, the dominant decay 
modes of the charged and neutral winos will be to $\nutil_{\tau}$ 
accompanied in the first case by a charged lepton and in the second by 
a neutrino. 
There is substantial $\ttil_{L,R}$ mixing and consequently radiative 
corrections raise $m_h$ above the 
current bound $m_h > 114 \GeV$\ref\hbound{LEP Working Group for 
Higgs boson searches, hep-ex/0107029}\
(we have included explicit radiative corrections other than 
leading logarithm effects in the 
calculation of $m_h$ only). 
Notice that $\etil_L,\mutil_L$ are 
quite light; in the so-called mAMSB model (where the slepton mass problem 
is resolved by adding a common $(\hbox{mass})^2$ to the scalars) this 
would be disfavoured due to the existence of charge-breaking extrema 
of the scalar potential\DattaER. 
We will investigate this possibility in our context 
elsewhere. 

We must consider the issue of  scalar-mediated FCNCs, which at first sight 
would appear to pose a real problem, because of our generation-dependent 
$U'_1$ charges. The AMSB
contributions to the scalar masses are diagonalised to a good approximation 
when we transform to
the  fermion mass-diagonal basis; as in fact are the FI contributions to
the LH squarks  and  sleptons, because of the universal 
doublet $U'_1$ charges. 
Thus the main source of  supersymmetric 
FCNC effects is  potentially from the RH squarks and sleptons.   
In the case of the squarks, these effects can be reduced by increasing
the gravitino mass $m_0$, which determines the scale of the AMSB 
contributions. However in the case of the sleptons, because of the 
crucial role of the FI terms, it is generally the case that 
some of the sleptons are comparatively light. 
In the sample spectrum (Eq.~\spect) note in particular that 
$m_{\etil_R} = 117\GeV$.  
As we indicated above, what in fact saves us from trouble is the fact
that  our choice of texture matrices means that the charged lepton
masses are   diagonalised by  transforming (to a good approximation) the
{\it LH fields only}.  

Consider, for example, 
the contribution to $\mu\to e\gamma$ 
from the neutralino/RH charged slepton loop. 
Because this contribution to the branching 
ratio is 
suppressed compared to the  chargino/LH sneutrino 
loop we are able to 
tolerate a larger 
amount of flavour mixing than when this mixing occurs in the LH sector.  
We find, in fact, that we typically obtain 
\eqn\mixing{
\delta^{RR}_{e\mu} = \frakk{m^2_{\etil_R\mutil_R}}{m^2_{\etil_R}} 
\sim 10^{-2}}
and that this leads to sufficient suppression of the  
branching ratio for $\mu\to e\gamma$  
for the kind of spectrum shown in Eq.~\spect.

The correlation between  the supersymmetric contributions to
$(g-2)_{\mu}$ and  $\mu\to e\gamma$ that has  been discussed by a number
of authors\MurakamiHK
\ref\gminustwo{ 
M.~Graesser and S.~Thomas, hep-ph/0104254\semi
Z.~Chacko and G.~D.~Kribs,
\prd 64  (2001) 075015\semi
J.~Hisano and K.~Tobe,
\plb 510 (2001) 197}
is weakened here because   the former is dominated by the
chargino/LH sneutrino   loop. For choices of $\zeta_1,\zeta_2$ such that
$\mutil_L, \etil_L$ are light compared to $\mutil_R, \etil_R$ 
it is possible to 
obtain within our framework a supersymmetric contribution to 
$(g-2)_{\mu}$ sufficient to explain the 
(now reduced
\ref\KnechtQF{
M.~Knecht and A.~Nyffeler, hep-ph/0111058}) discrepancy 
between the Standard Model and experiment\ref\brown{H.N.~Brown {\it et al}, 
\prl 86 (2001) 2227} while being consistent with the limit on 
$\mu\to e\gamma$.

We now turn to the $\Delta\neq 0$ case. 
The Higgs 
$\mu$ term is then forbidden by the $U'_1$ symmetry, but 
if $|\Delta|$ is sufficiently large, then we can imagine 
generating the $\mu$-term via the FN-mechanism\ref\GogoladzeKJ{ I.~Gogoladze 
and A.~Perez-Lorenzana,  hep-ph/0112034}, i.e. via an interaction 
of the form $M_U H_1H_2(\frakk{\theta_H}{M_U})^{n_H}$ where 
$n_H\approx 16$. The resulting sparticle spectrum is broadly similar 
to that considered above in the $\Delta = 0$ case, and   neutrino 
masses can be introduced in a similar way.  Depending on the $\theta$-charge 
assignments, 
it may be that whereas proton decay is adequately suppressed, 
lepton number-violating 
$R$-parity-violating operators are generated by the texture mechanism causing 
decay of the LSP. 

To summarise: with a generation-dependent $U'_1$  charge assignment we 
are able to extend our previous FI-based solution to the AMSB 
tachyonic slepton problem to accommodate neutrino masses. 
We achieve natural suppression of leptonic FCNCs via a combination of 
universal doublet $U'_1$ charges and textures which are diagonalised 
primarily by LH transformations. We will consider in more 
detail elsewhere the constraints placed on our general 
framework by experimental limits on sparticle masses, and other issues 
such as vacuum stability and CP-violation.

\bigskip\centerline{{\bf Acknowledgements}}

DRTJ and RW were  supported by a PPARC Senior Fellowship and  
a PPARC Graduate Studentship respectively. DRTJ is grateful to the 
Institute of Theoretical Physics, at SUNY Stony Brook
for financial support, and its members  
for hospitality, while part of this work was done.  We thank Barry King, 
John McDonald and Apostolos Pilaftsis for conversations.

\listrefs

\end